\documentclass[aps,prb,twocolumn,nopacs,floats,superscriptaddress]{revtex4-2}
\usepackage{lineno}       % line numbers
\usepackage{hyperref}     % hyperlinks
\usepackage{amsmath,amssymb} % math symbols
\usepackage{graphicx}     % figures
\usepackage{float}        % figure placement
\usepackage{soul} %text highlighting
\usepackage{xcolor}
\sethlcolor{yellow} % Set custom highlight color

\begin{document}

\title{Ground state correlations in the one-dimensional Fermi one-component plasma}
\date{March 2026}

\author{Massimo Boninsegni}
\affiliation{Department of Physics, University of Alberta, Edmonton, Alberta, Canada T6G 2H5}
\affiliation{Phenikaa Institute for Advanced Study, Phenikaa University, Nguyen Trac Street, Duong Noi Ward, Hanoi, Vietnam}

\begin{abstract}
Structural and dynamic correlations in the ground state of the one-dimensional Fermi one-component plasma are studied by Quantum Monte Carlo simulations. Results are presented for the pair correlation function, static structure factor, the one-particle density matrix and the momentum distribution, for the cases of full, partial and no polarization.    {Within the precision of the calculation, the results conform to the fundamental scaling prediction of the Tomonaga-Luttinger Liquid Theory}. Evidence is reported of density correlations slowly decaying with distance, with the concurrent emergence of quasi-crystalline order even in the weakly correlated regime. Effects of quantum statistics in the momentum distribution are discussed.
\end{abstract}

%\begin{keyword}
%Superfluidity \sep Quantum Monte Carlo simulations %\sep Bose one-component plasma \sep Berezinskii-%Kosterlitz-Thouless transition
%\end{keyword}
\date{\today}
%\end{frontmatter}
\maketitle
\section{Introduction}
A system of spin-1/2 charged particles moving in the presence of a uniform neutralizing background of opposite charge is possibly the most extensively studied model in theoretical condensed matter physics. Known as {\em jellium}, electron gas, or, as we shall refer to it here, one-component plasma (OCP), this model is regarded as foundational in the understanding of electronic properties of metals, or the effects of strong electron correlations, collective phenomena such as plasmons, screening and Friedel oscillations. As well,
it is the basic building block of just about any computational framework aimed at investigating real materials, such as Density Functional Theory \cite{Giuliani2005}.
\\ \indent 
Interest in the one-dimensional Fermi OCP dates back decades \cite{Devreese1979}; while in the past the study of many-body systems confined to one spatial dimension might have been regarded as of merely conceptual, academic relevance, there now exist experimental systems in which electron motion is confined to one dimension, for practical purposes \cite{Bockrath1999,Ishii2003,Chang2003,Schafer2008,Huang2001,Moritz2005}. Thus, a meaningful comparison of theory and experiment is not out of the question.
\\ \indent
From the theoretical standpoint, one spatial dimension occupies a special place within the many-body problem, as accepted theoretical frameworks that broadly describe general behavior in higher dimensions (e.g., Landau Fermi Liquid Theory, LFLT) no longer hold.    {For example, in one dimension effects of quantum statistics are blurred} \cite{Girardeau1960,Yukalov2005}   {; for Fermi systems, this
%MB July 14
%almost entirely suppressed by repulsive interactions among particles 
can result in the disappearance of a sharp Fermi surface, hence of the conventional particle-hole excitations that are a cornerstone of the LFLT.}
%MB causes 
\\ \indent
A single unified description of one-dimensional (1D) Fermi and Bose systems is provided by the Tomonaga-Luttinger Liquid theory (TLLT), which affords a number of remarkably precise predictions \cite{Tomonaga1950,Luttinger1963,Haldane1981}. In order to test such predictions, a significant effort over the past two decades has been devoted to the identification and realization in the laboratory of (quasi) 1D many-body systems. 
The 1D Fermi OCP is particularly interesting because it does {\em not} conform to the standard TLLT, due to the long-ranged nature of the interaction, whose  main consequence is the absence of conventional low-lying elementary excitations with a linear dependence of the energy on the wave vector $q$, in the long-wavelength limit. Rather, the dispersion curve takes the form $\omega_q\sim q\sqrt{{\rm ln}(1/q)}$ \cite{Giamarchi2003}, i.e., there is no  well-defined speed $c$ associated with the elementary excitations and therefore no characteristic parameter $K=c/v_F$ (where $v_F$ is the Fermi velocity), which governs the physics of conventional Luttinger liquids.
\\ \indent 
The 1D OCP has been predicted to display distinct behavior, e.g., an enhanced tendency to develop quasi-crystalline order even in the weakly interacting (high density) limit \cite{Schulz1993}. It seems that these predictions could be tested by Quantum Monte Carlo (QMC) simulations, which have played an important role in shaping our current understanding of the physics of the Fermi OCP, in three \cite{Ceperley1980,Moroni1992,Ortiz1994,Spink2013,Dornheim2016,Dornheim2018,Dornheim2018b,Azadi2021} 
and two \cite{Tanatar1989,Kwon1996,Drummond2009} dimensions. In particular, deviations of numerical data from the most important scaling predictions of the TLLT may be observed.
\\ \indent
Somewhat surprisingly, however, relatively little has been published so far for the 1D case, despite the fact that in one dimension no sign problem arises, and virtually exact numerical estimates can be obtained for all the physically relevant quantities, including the momentum distribution, embodying all detectable signatures of quantum statistics. This is in stark contrast with two and three dimensions, where QMC is affected by the well-known ``sign'' problem, which limits its applicability and generally requires some workarounds, typically involving uncontrolled approximations (e.g., the fixed-node one).  
\\ \indent
The few QMC studies of the 1D Fermi OCP have been based on variational \cite{Casula2005,Ashokan2018} and Diffusion Monte Carlo \cite{Lee2011} simulations    {(the latter for a fully polarized system only)}. Although they have provided considerable insight, these calculations are inherently biased by the specific choice of trial wave function utilized. In DMC some of the bias can be removed; for the ground state energy, in principle an exact result can be obtained, as long as a careful (and often costly) extrapolation is performed with respect to the number of random walkers and projection time \cite{Nemec2010,Boninsegni2012,Boninsegni2001}. On the other hand, for other observables the bias arising from the trial wave function is never completely removed, especially if, as it is almost invariably the case, the so-called ``mixed estimators'' are used. It seems to make sense, therefore, to carry out a separate, independent QMC study of the ground state of the 1D Fermi OCP making use of a different technique, one not affected by any specific choice of trial wave function, or any other {\em a priori} input.
\\ \indent
We illustrate in this paper the results of a theoretical investigation of the 1D Fermi OCP based on QMC simulations. We make use of a finite temperature technique, and attempt to extract ground state (i.e., zero-temperature) structural and dynamic correlations in the thermodynamic limit by collecting results at sufficiently low temperature. 
The standard procedure to achieve that goal is that of collapsing data for varying temperature $T$ and systems size $L$, taking advantage of an exact scaling property  derivable within the TLLT \cite{Delmaestro2011,Boninsegni2013,Boninsegni2025}, according to which, in the limit $L\to\infty$ and $T\to 0$, all physical averages should depend only on the product $LT$. Because the standard TLLT is not applicable, deviations from this behavior are in principle expected. However, at least to our knowledge it is not clear how
quantitatively significant and/or readily observable they are in practice, and, conversely, whether the standard TLLT theory may still yield useful indications. We therefore cautiously adopt the product $LT$ to guide our analysis of our numerical data.  Our main purpose is to obtain possibly new insight by providing estimates for quantities for which none are yet available; for example, while the study of Ref. \cite{Lee2011} is restricted to the fully polarized case, here we also consider the partially polarized and unpolarized cases, in order to expose how the momentum distribution reflects the state of polarization of the system. 
\\ \indent
This paper is organized as follows: in section \ref{meth} we introduce the model and describe our methodology, present our results in sec. \ref{res} and outline our conclusions in Sec. \ref{concl}.

\section{Methodology} \label{meth}
We consider an assembly of $N$ identical spin-1/2 charged particles moving in a one-dimensional box of length $L$ with periodic boundary conditions. Let $\zeta$ be the net system polarization, i.e., a fraction $(1+\zeta)/2$ of the $N$ particles have positive spin projection. Particles interact via the standard pairwise, repulsive Coulomb interaction, $\propto 1/r$; moreover, a uniform background of charge of sign opposite to that of the particles is included, so as to ensure overall electric neutrality. \\ \indent 
We use    {standard} (Hartree) atomic units and characterize the density of the system through the average interparticle distance, conventionally expressed by the (dimensionless) parameter $r_s\equiv L/(2Na)$, $a$ being the unit    {of length} (Bohr radius). \\ \indent 
The resulting quantum-mechanical many-body Hamiltonian of the system is defined as follows:
\begin{equation}\label{ham}
    \hat H=-\frac{1}{2}\sum_i\frac{\partial^2}{\partial x_i^2}+\sum_{i < j} V(x_{ij}) +\frac{N}{2}V_M
\end{equation}
where $x_i$ is the position of the $i$th particle, $x_{ij}\equiv |x_i-x_j|$,  and 
\begin{equation}\label{ewald}
V(x_{ij}) = \sum_{n=-\infty}^{+\infty}\biggl ( \frac{1}{|x_{ij}+nL|}-\frac{1}{L}\int_{-L/2}^{L/2}\  \frac{dy}{|x_{ij}-y+nL|}\biggr )
\end{equation}
is the (Ewald) potential describing the Coulomb interaction of an  electron with (all the periodically replicated images of) another electron whose closest image is at a distance $|x|$, and with a fraction worth 1/$N$ of the uniform positive background. This is the interaction utilized in Ref. \cite{Lee2011}. $V_M$ is the so-called Madelung energy, i.e., the interaction of every particle with the background and with all its periodic images.
An accurate numerical evaluation of (\ref{ewald}) can be achieved through the approximation described in Ref. \cite{Saunders1994} (Eq. 4.8).
In practice, as is customarily done, we tabulate (\ref{ewald}) over sufficiently fine a grid of distances extending up to $L/2$, and compute the interaction on the fly by cubic spline interpolation.
\\ \indent
We computed thermodynamic properties of the system described by Eq. \ref{ham} by means of QMC simulations at finite temperature ($T$), based on the canonical variant \cite{Mezzacapo2006,Mezzacapo2006b} of the continuous-space Worm Algorithm \cite{Boninsegni2006,Boninsegni2006b}. We obtained results in the density range $1\le r_s\le 20.$. Details of the simulation are  standard; we used a short-time propagator which is accurate to fourth order in the time step $\tau$ \cite{Jiang2001,Boninsegni2005}. We carried out numerical extrapolation of the estimates to the $\tau\to 0$ limit, and generally observed convergence of the thermal averages for a value of $\tau\approx {10^{-3}}\ {r_s}^{2}$ Ha$^{-1}$ for all quantities of interest here. These include the energy per particle, the pair correlation function and the one-body density matrix. Unlike in DMC, these quantities can be computed using unbiased estimators, not affected by a specific choice of trial ground state wave function.
\\ \indent
As mentioned at the outset, a QMC simulation of this system is not affected by the infamous ``sign'' problem that plagues    {QMC} studies of Fermi systems in two and three dimensions.    {This is because in the thermodynamic limit only the identity permutation contributes to the partition function. Thus, the equilibrium properties are for the most part those of a system of}    {\em distinguishable}    {particles. However, off-diagonal correlations such as the one-body density matrix  (and the related momentum distribution) display distinct signatures of quantum statistics, and therefore are quantitatively different, e.g., in polarized and unpolarized systems (the non-interacting Fermi gas providing the simplest example)}\cite{Girardeau1960}.
\\ \indent     {For a purely repulsive Coulomb system, the interaction acts to prevent particles from approaching each other, regardless of their relative spin orientation; moreover, its long-range character enhances Wigner crystallization, and the conventional wisdom is that effects of quantum statistics are suppressed, and in the low density (strongly interacting) limit thermal averages of energetics and structural properties become independent of the polarization} $\zeta$.     {It is an interesting question to what a degree, and in what range of density, effects of Fermi statistics are observable in the momentum distribution.}
\section{Results}\label{res}
We begin by discussing the energetics, not because it is is of particular interest in and of itself but because it allows one to make a quantitative, direct comparison with other calculations, in order to check for consistency and validate the methodology and the results presented. As mentioned above, we aim at extracting ground state ($T=0$) results in the thermodynamic ($L\to\infty$) limit. 
As mentioned above, the TLLT establishes that in the limit $L\to\infty, T\to 0$ the values of all physical quantities must depend only on the product $LT$ (obviously, since $L$ is the system size and we work at constant density, doubling $L$ implies doubling $N$) \cite{Delmaestro2011,Boninsegni2013}. However, the system of interest here ought  {\em not} to conform to that, as the standard TLLT does not apply, i.e., corrections to such scaling must emerge in the low temperature limit, though they may not be easily observable in a numerical simulation. 
\begin{figure}
\includegraphics[width=0.47\textwidth]{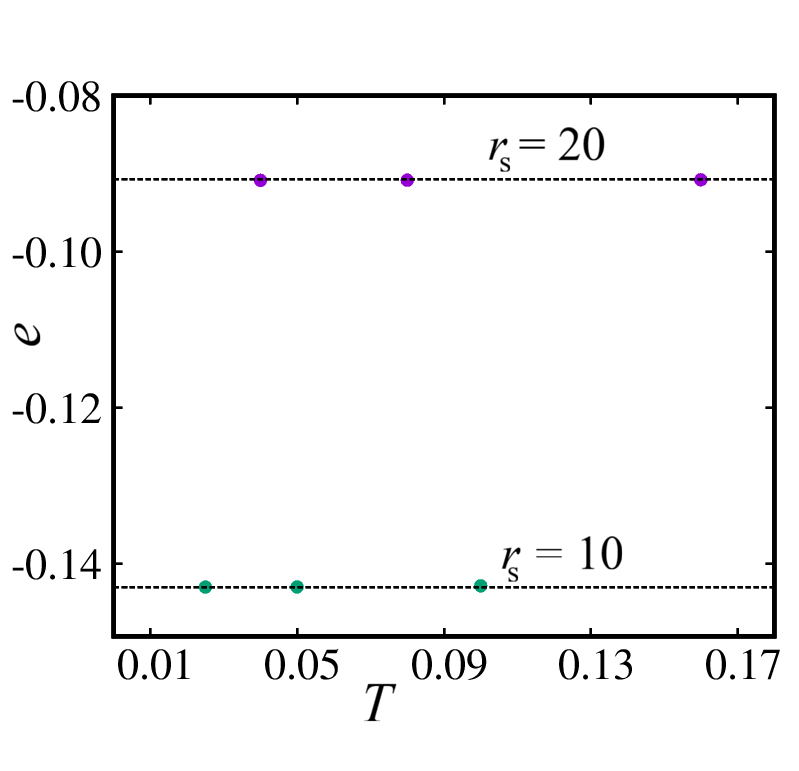}
\caption{Energy per particle as a function of temperature (both quantities in Hartree) for a fully polarized system of electrons and two different values of $r_s$. The numbers of particles are 20 (highest temperature), 40 and 80 (lowest temperature). Horizontal dashed lines are fits to the data. Statistical errors are smaller than symbol sizes. }\label {cazzupo}
\end{figure}
\\ \indent
Fig. \ref{cazzupo} shows the computed energy per particle as a function of temperature for a fully polarized system of electrons, with $r_s=10$ and 20. The values are obtained with different numbers of particles, specifically $N=20$ (highest temperatures), 40 and 80 (lowest temperature), so that the product $LT$  is constant.
The dashed horizontal lines are fits to the data, i.e., no difference can be detected, within the statistical errors of our calculation, among results obtained at different temperatures. We observed this kind of behavior at low $T$ for all the values of $r_s$ considered here. We therefore report as ground state energy estimates the results arrived at through this extrapolation procedure; they are listed in Table \ref{energyresults}, where they are compared to those of Ref. \cite{Lee2011}, which are obtained by extrapolating to the thermodynamic limit estimates for systems of size up to $N=100$ particles.
\begin{table}[htbp] % Placement options: h=here, t=top, b=bottom, p=page of floats
    \centering % Center the table
    \caption{Estimated ground state energy per electron    {(in Hartree)} at different values of $r_s$, obtained as illustrated in Fig. \ref{cazzupo}. Statistical errors (in parentheses) are on the last digit.} 
    \label{energyresults} % Label for referencing
    \begin{tabular}{c|c|c} % Column alignment: l=left, c=center, r=right
        $r_s$ & This work & Ref. \cite{Lee2011} \\    ine
        20. & $-0.09078(3)$ & $-0.090777768(2)$ \\ 
        10. & $-0.14293(3)$ & $-0.142869097(9)$ \\ 
        5. & $-0.2027(2)$ & $-0.20393235(2)$ \\ 
        2. & $-0.2080(9)$ &$-020620084(7)$ \\ 
        1. & 0.1563(8) & 0.1541886(2) \\    ine
    \end{tabular}
\end{table}
\\ \indent
The two independent calculations provide results consistent with each other, although the DMC ones are affected by a much smaller statistical uncertainty. 
We take this as an indication of the correctness of our calculation; we did not seek to achieve the same level of precision in evaluating this quantity, because it was not the main focus of our investigation, which was rather the study of structural and dynamic correlations.
\begin{figure}[h]
\includegraphics[width=0.47\textwidth]{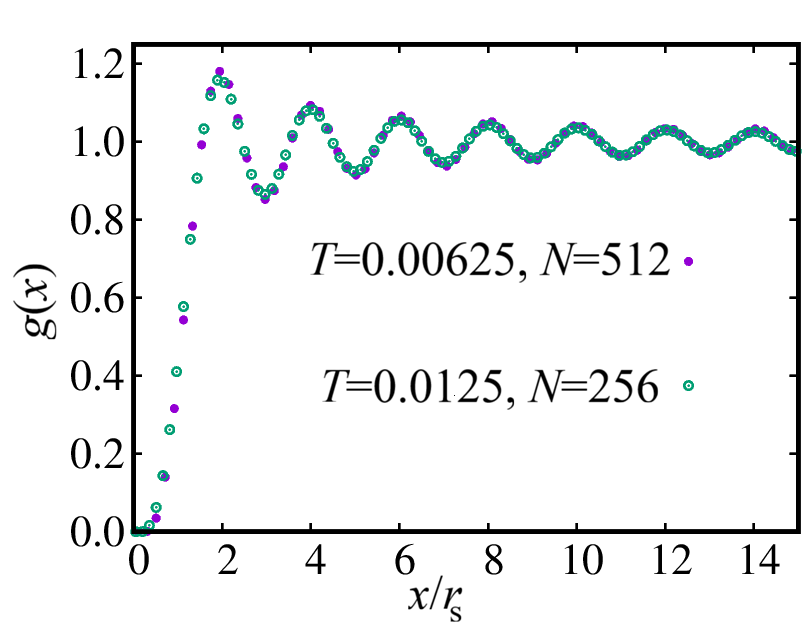}
\caption{Pair correlation function $g(x)$ computed for a fully polarized ($\zeta=1$) system with $r_s=2$ at two temperatures (in Ha) and with system sizes so that the product $LT$ remains constant. The data are observed to collapse (statistical errors are of the order of the symbol size).}\label {collapse}
\end{figure}
\\ \indent
Fig. \ref {collapse} displays the pair correlation function $g(x)$ for a fully polarized system with $r_s=2$, plotted as a function of $x/r_s$. Within the statistical uncertainties of our calculation, collapse of the data is observed; just like for the energy, we find that to be the case for all values of $r_s$ considered here.
\begin{figure}[h]
\includegraphics[width=0.47\textwidth]{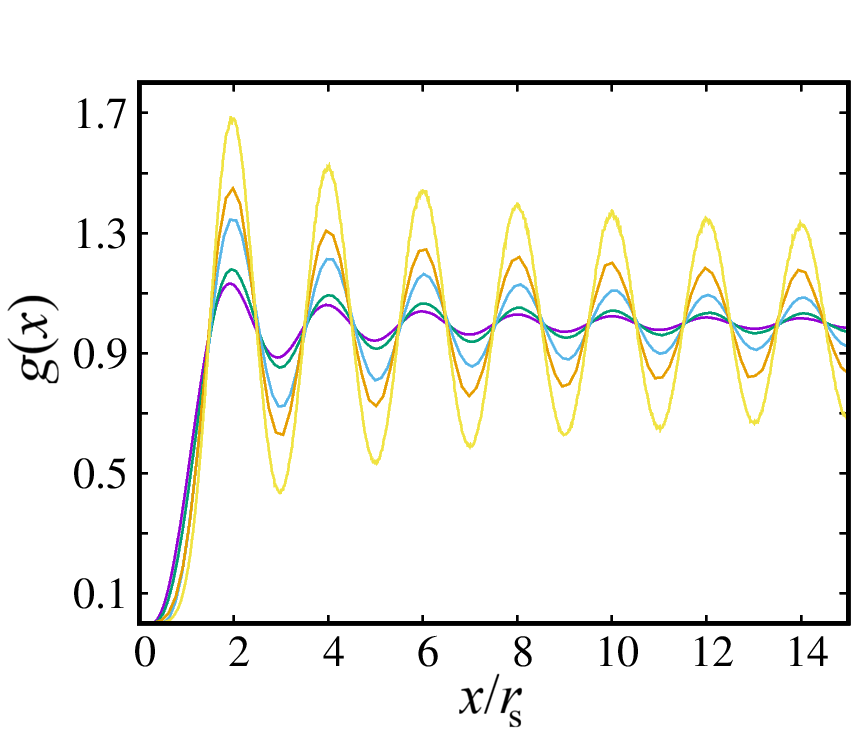}
\caption{Ground state pair correlation function $g(x)$ computed for a polarized system with $r_s=1, 2, 5, 10, 20$. Curves with higher peaks correspond to greater values of $r_s$.}\label {gofr}
\end{figure}
Fig. \ref{gofr} shows the so determined ground state pair correlation functions for a fully polarized (i.e., $\zeta=1$) system of varying $r_s$, specifically $r_s=1,2,5,10,20$. The curves with higher peaks correspond to greater values of $r_s$ (i.e., lower density). There is satisfactory quantitative agreement with the results offered in Ref. \cite{Lee2011}. In general, the height of the first peak is merely $\sim$1-2\% higher in this calculation, but altogether the comparison seems to provide a strong validation for both studies. Such an agreement is noteworthy, as the calculation of Ref. \cite{Lee2011} made use of the ``mixed'' estimators, which in principle are biased by the trial wave function adopted. The close numerical agreement with our (unbiased) results points to the high accuracy of the trial wave function utilized in Ref. \cite{Lee2011}.
\\ \indent
The next quantity that we consider is the static structure factor $S(q)$. This can in principle be computed from the $g(x)$ by carrying out a Fourier transform, on extending the $g(x)$ outside the simulation cell, not to incur loss of accuracy. This procedure becomes in our view problematic and ambiguous for $r_s$ even moderately large (i.e., $\gtrsim 2$), as the $g(x)$ features oscillations that die off quite slowly with $x$. Thus, in this study we opted to compute the static structure factor directly.
\begin{figure}[h]
\includegraphics[width=0.48\textwidth]{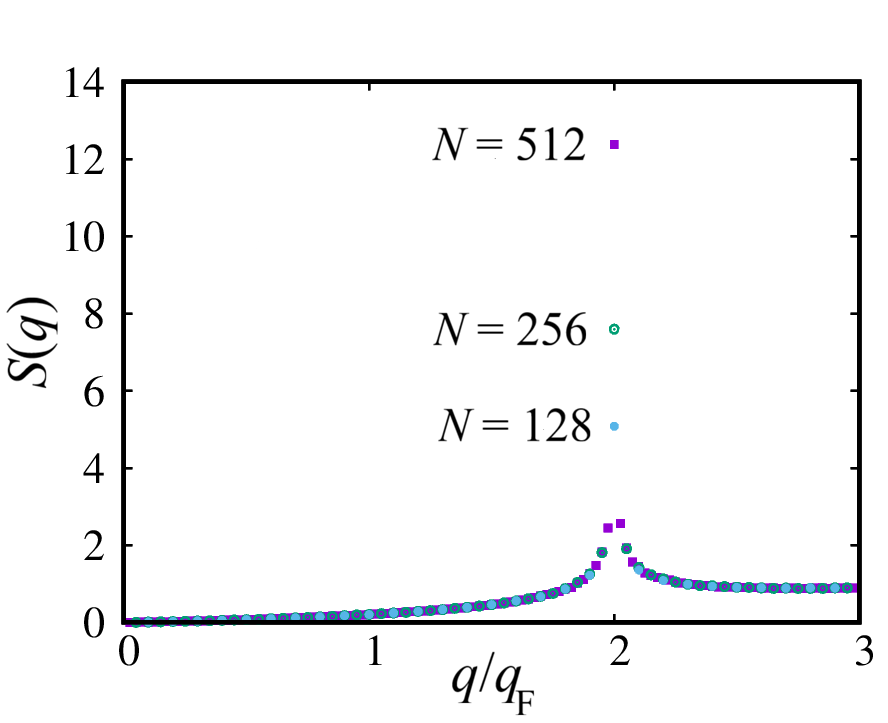}
\includegraphics[width=0.48\textwidth]{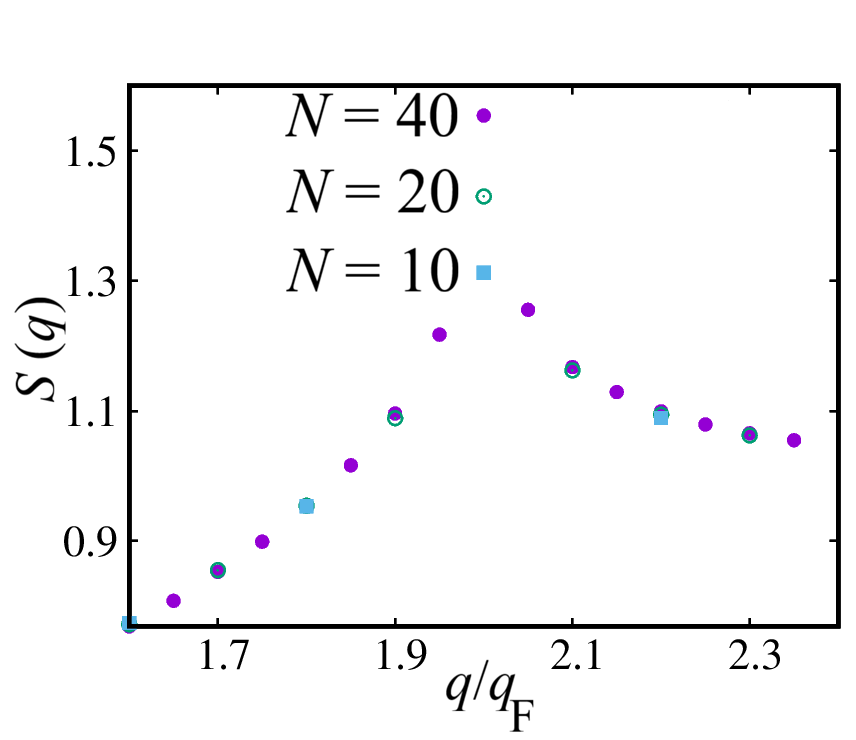}
\caption{Static structure factor $S(q)$ computed for a fully polarized ($\zeta=1$) system with $r_s=20$ (top) and $r_s=2$ (bottom). Data are plotted as a function of $q/q_F$, where $q_F=\pi/(2r_s)$. Statistical errors are smaller than symbol size. The data shown pertain to three different system sizes and temperatures. The lowest temperature is $3.125 \times 10^{-2}$ Ha, and the product $LT$ is the same for all systems.}\label {sq20}
\end{figure}
\\ \indent
The result for a fully polarized electron gas is  shown in Fig. \ref {sq20} for $r_s=20$ (top) and $r_s=2$ (bottom); the static structure factor $S(q)$ is plotted against $q/q_F$, where $q_F=\pi/(2r_s)$ is the Fermi wave vector. 
A similar collapse of data observed for the pair correlation function takes place for the static structure factor, {\em except} in the vicinity of the Fermi wave vector $q_F$, where the static structure factor diverges in the thermodynamic limit, as shown for instance in the figure. 
It is known that no true long-range order can exist in one dimension. However, a quasi-crystalline phase characterized by divergent (Bragg) peaks in the static structure factor at wave vectors multiple of $q_F$ emerges as the density of the system is lowered and the potential energy dominates over the kinetic energy. 
\\ \indent
In two and three dimensions a first-order quantum phase transition occurs at zero temperature from a crystal to a fluid phase as the density is increased. On the other hand, the long-range character of the Coulomb interaction is predicted to stabilize a low temperature (quasi) crystalline phase at arbitrarily high density \cite{Schulz1993}. The results obtained in this work support such a contention. As $r_s$ is lowered, the divergence with system size of the peak of the static structure factor at $q=2q_F$ occurs more slowly, but remains clearly visible, as shown for example in Fig. \ref{sq20} for $r_s=2$. Another intriguing prediction is the emergence of antiferromagnetic order in the unpolarized system, signaled by a peak at $q=4q_F$ \cite{Giamarchi2003}; this aspect was not investigated in this work, however.
\begin{figure}[h]
\includegraphics[width=0.48\textwidth]{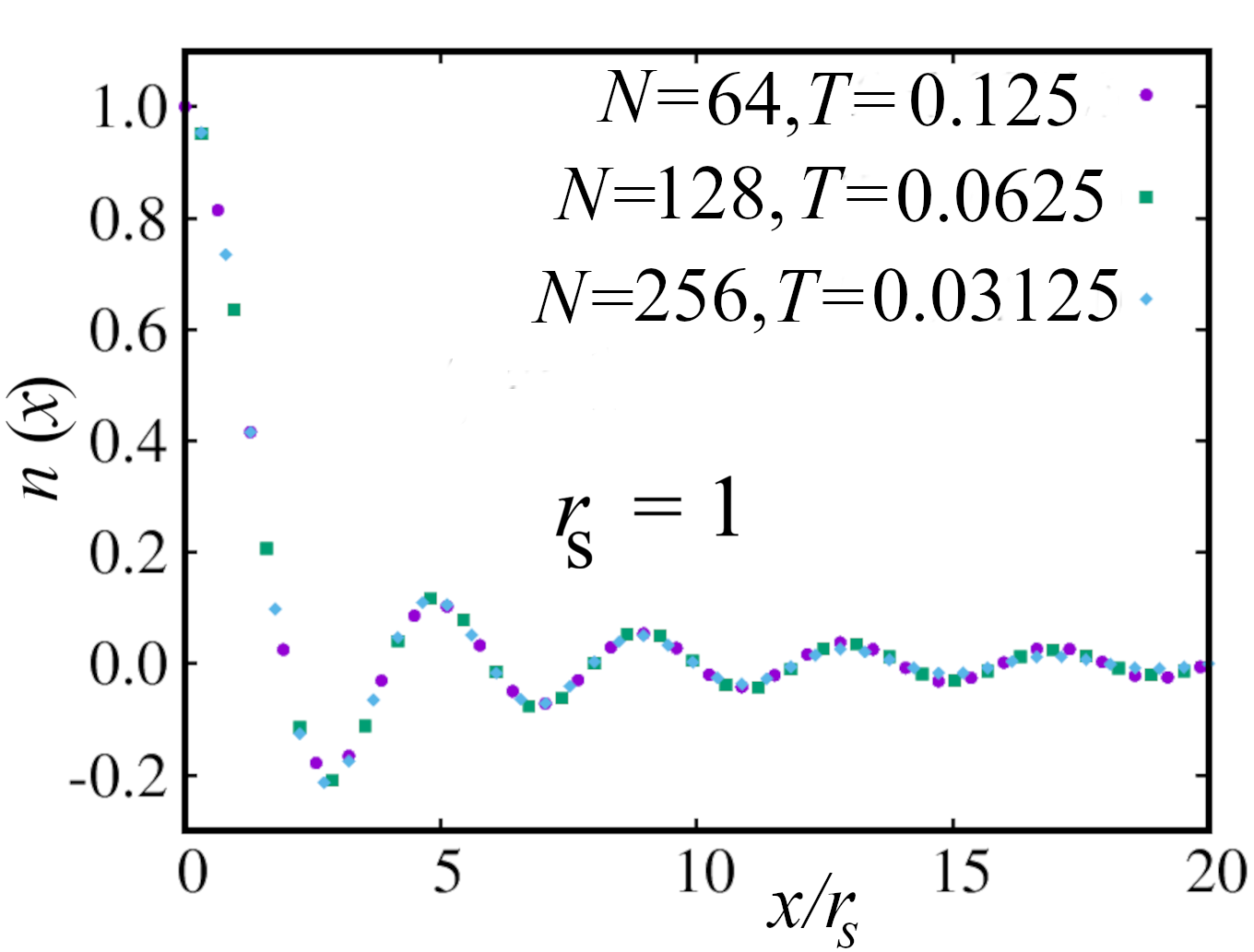}
\includegraphics[width=0.48\textwidth]{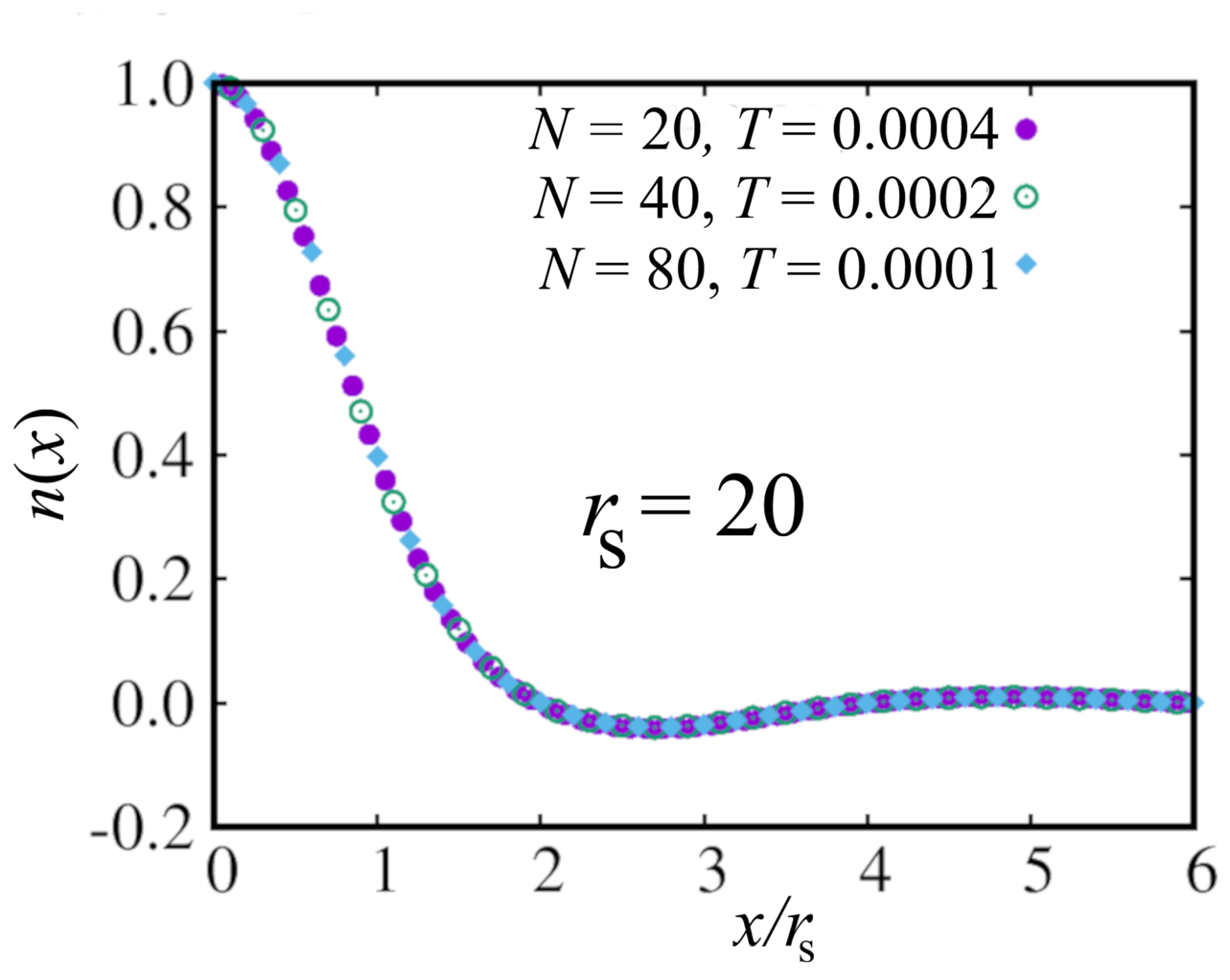}
\caption{Ground state one-particle density matrix $n(x)$ computed for a fully polarized ($\zeta=1$) system with $r_s=1$ (top) and $r_s=20$ (bottom). Data are plotted as a function of $x/r_s$. Statistical errors are smaller than symbol size. The data shown pertain to three different system sizes and temperatures.}\label {Obdm}
\end{figure}
\begin{figure}[h]
\includegraphics[width=0.48\textwidth]{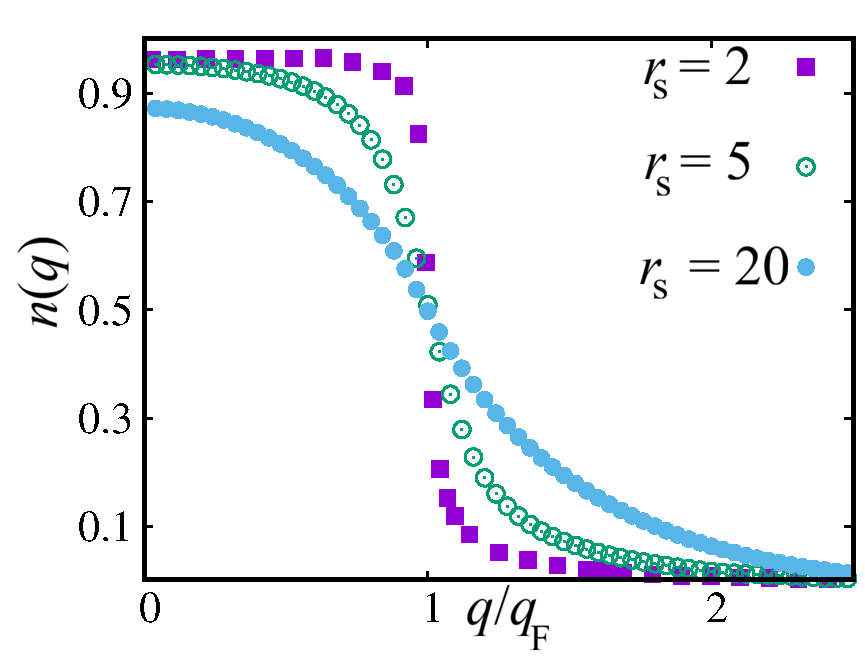}
\caption{Ground state momentum distribution $n(q)$ computed for a fully polarized ($\zeta=1$) system with $r_s=2, 5, 20$. Data are plotted as a function of $q/q_F$, with $q_F=\pi/(2r_s)$. Statistical errors are smaller than symbol size.}\label {nofq}
\end{figure}
\\ \indent
Given the preponderance of crystalline order in this system, it is interesting to see to what a degree that is reflected in the momentum distribution, which we obtain as Fourier transform of the one-body density matrix $n(x)$, for which results are shown in Fig. \ref{Obdm} for a fully polarized system.
In this case too, collapse of the data is impressive in the whole range of $r_s$ explored; thus, just like for the pair correlation function, this provides a justification for considering the results as ground state estimates, and as such are being reported. At high density ($r_s=1$), $n(x)$ displays the characteristic (Friedel) oscillations that are typical of weakly interacting Fermi systems. As the density is lowered, the oscillations become progressively less pronounced, even though even at $r_s=20$ $n(x)$ goes negative for $x \gtrsim 2\ r_s$. 
\\ \indent 
The momentum distribution for a fully polarized system is shown in Fig. \ref{nofq}. The qualitative change as the density is lowered is clear, with the $n(q)$ displaying a sharp step at $q=q_F$ for $r_s=2$, becoming then smoother and with a slowly decaying tail at high $q$ for $r_s=5, 20$. Altogether, the results are in satisfactory agreement with those of Ref. \cite{Lee2011}. Qualitatively, the behavior of the momentum distribution as a function of $r_s$ resembles that in higher dimensions. At high density ($r_s\to 0$) $n(q)$ features the sharp demarcation between occupied ($q\lesssim q_F$) and unoccupied ($g\gtrsim q_F$) single-particle states hat characterizes the ground state of the non-interacting system. On the other hand, as the density is lowered and the effect of interactions becomes increasingly important, and with it the occupation of high-momentum states. In this limit, $n(q)$ is smooth and slowly decaying. The noteworthy difference is that in higher dimensions interactions cause crystallization at low density, whereas in 1D  a Wigner crystal forms at all densities, a fact of which the momentum distribution (at last computed as illustrated here) gives no hint.
\\ \indent
It is interesting to examine the effect of polarization on the $n(q)$.
\begin{figure}[h]
\includegraphics[width=0.48\textwidth]{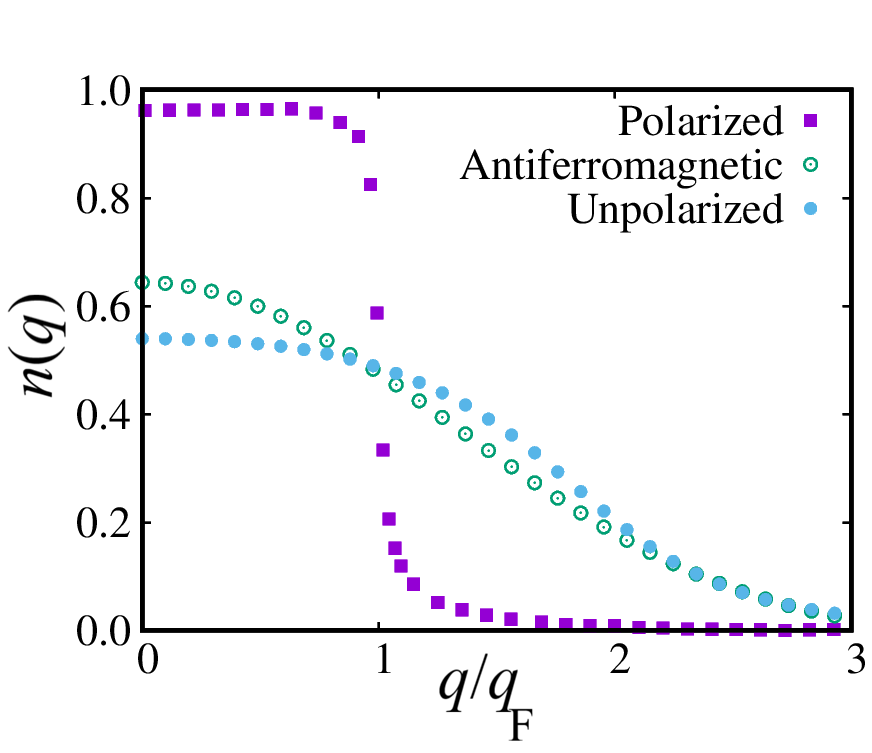}
\includegraphics[width=0.48\textwidth]{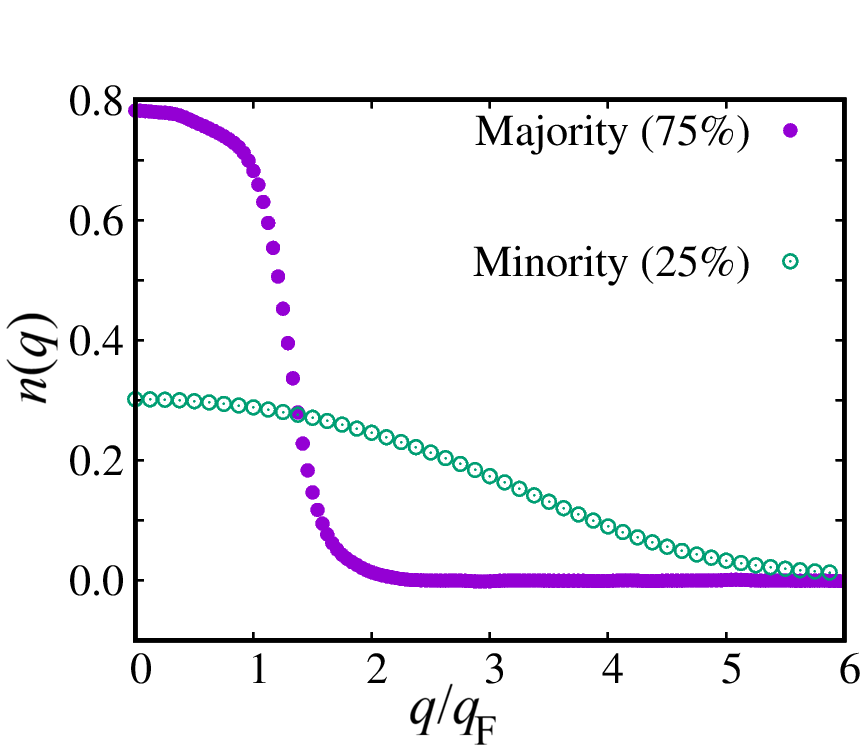}
\caption{{\em Top}: Ground state momentum distribution $n(q)$ computed for a fully polarized ($\zeta=1$, boxes) and unpolarized ($\zeta=0$, filled circles)  system with $r_s=2$. Results are plotted as a function of $q/q_F$, where $q_F=\pi/(2r_s)$ [$q_F=\pi/(4r_s)]$ for $\zeta=1$ [$\zeta=0$]. Also shown is the result for antiferromagnetic spin alignment (open circles). {\em Bottom}: Momentum distribution fothe majority and minority spin components, for a system with $\zeta=0.5$. 
Statistical errors are smaller than symbol size.}\label {pnofq}
\end{figure}
Fig. \ref{pnofq} (top) shows the ground state momentum distribution for an unpolarized ($\zeta=0$) system with $r_s=2$. Also shown for comparison are the results for full polarization (i.e., $\zeta=1$), as well as for an unpolarized system in which electrons are antiferromagnetically aligned, which corresponds to particles being {\em distinguishable} \cite{Boninsegni2025}. Simulations of unpolarized systems were carried out by averaging over a set (typically 10) of different realizations, each featuring a different random spin alignment. The results are plotted as a function of $q/q_F$, where $q_F=\pi(2r_s)^{-1}$ for the polarized system, and $q=\pi(4r_s)^{-1}$ for the unpolarized one. The momentum distribution of the unpolarized system displays little or no sign of quantum statistics, as expected. This is most pronounced for the case where the spin alignment is fully antiferromagnetic, i.e., each electron has nearest neighbors with opposite spin projection. In this case, as a results of the short-distance repulsion of the Coulomb interaction, particles can be regarded as distinguishable, and the momentum distribution is nearly classical.
\\ \indent The bottom part of Fig. \ref{pnofq} shows instead the momentum distribution for the majority and minority spin component for a system with partial ($\zeta=0.5$) polarization ($r_s=2$). Here too, the results are plotted versus $q/q_F$, where $q_F$ reflects the different density of the two components. The comparison highlights the importance of quantum statistics for the majority components, as the majority component electrons undergo exchanges with identical nearest neighbors more frequently than their minority component counterpart.
\section{Conclusions}\label{concl}
We have carried out an extensive study, based on computer simulation, of the ground state properties of the one-dimensional electron gas (Fermi one-component plasma), in the range of mean interparticle separation $1 \le r_s\le 20$. This system is known not to conform to the standard TLL theory \cite{Giamarchi2003}, and it seems worthwhile to investigate its low temperature properties based on an unbiased numerical QMC technique which in one dimension does not suffer from any sign instability. To our knowledge, there is only one previous comparable study, based on a different (ground state) method \cite{Lee2011}, which however only considered the fully polarized case.
\\ \indent
Within the precision of our calculation we could not detect any significant deviation of our results from the main scaling prediction of the TLL theory; specifically, we generally observed collapse of the data in the $L\to\infty, T\to 0$ limit. The results quoted herein as ``ground state'' reflect this observed behavior; they are generally in excellent agreement with those of Ref. \cite{Lee2011}, as far as energetic and structural properties are concerned.
\\ \indent
Our results are consistent with Wigner crystallization for any value of $r_s$, as shown by a visible peak in the static structure factor, whose height is a monotonically increasing function of system size. Interestingly, distinct signatures of quantum statistics remain, as shown by the momentum distribution computed as a function of system polarization. In other words, quantum-mechanical exchanges which are known to play a crucial role in shaping the phase diagram of many-body systems in higher dimensions \cite{Boninsegni2012b} but are suppressed in one dimension by the repulsive nature at short distance of the Coulomb interaction, nevertheless remain important, underscoring how differences between phases are blurred in one dimension.
\\ \indent
This work was supported by the Natural Sciences and Engineering Research Council of Canada, under grant RGPIN 2024-05664. The author is thankful to N. Drummond for useful comments and suggestions.
%\section*{References}
\bibliography{refs}

@book{Giuliani2005, 
    place={Cambridge}, 
    title={Quantum Theory of the Electron Liquid}, 
    publisher={Cambridge University Press}, 
    author={Giuliani, G. and Vignale, G.}, 
    doi = {10.1017/CBO9780511619915},
    year={2005}
}

@article{Ceperley1980,
  title = {Ground State of the Electron Gas by a Stochastic Method},
  author = {Ceperley, D. M. and Alder, B. J.},
  journal = {Phys. Rev. Lett.},
  volume = {45},
  issue = {7},
  pages = {566--569},
  numpages = {0},
  year = {1980},
  month = {Aug},
  publisher = {American Physical Society},
  doi = {10.1103/PhysRevLett.45.566}
}

@article{Ortiz1994,
  title = {Correlation energy, structure factor, radial distribution function, and momentum distribution of the spin-polarized uniform electron gas},
  author = {Ortiz, G. and Ballone, P.},
  journal = {Phys. Rev. B},
  volume = {50},
  issue = {3},
  pages = {1391--1405},
  numpages = {0},
  year = {1994},
  month = {Jul},
  publisher = {American Physical Society},
  doi = {10.1103/PhysRevB.50.1391}
}

@article{Dornheim2016,
  title = {Ab Initio Quantum Monte Carlo Simulation of the Warm Dense Electron Gas in the Thermodynamic Limit},
  author = {Dornheim, T. and Groth, S. and Sjostrom, T. and Malone, F. D. and Foulkes, W. M. C. and Bonitz, M.},
  journal = {Phys. Rev. Lett.},
  volume = {117},
  issue = {15},
  pages = {156403},
  numpages = {6},
  year = {2016},
  month = {Oct},
  publisher = {American Physical Society},
  doi = {10.1103/PhysRevLett.117.156403}
}

@article{Moroni1992,
  title = {Static response from quantum Monte Carlo calculations},
  author = {Moroni, S. and Ceperley, D. M. and Senatore, G.},
  journal = {Phys. Rev. Lett.},
  volume = {69},
  issue = {13},
  pages = {1837--1840},
  numpages = {0},
  year = {1992},
  month = {Sep},
  publisher = {American Physical Society},
  doi = {10.1103/PhysRevLett.69.1837}
}

@article{Azadi2021,
  title = {Quasiparticle Effective Mass of the Three-Dimensional Fermi Liquid by Quantum Monte Carlo},
  author = {Azadi, S. and Drummond, N. D. and Foulkes, W. M. C.},
  journal = {Phys. Rev. Lett.},
  volume = {127},
  issue = {8},
  pages = {086401},
  numpages = {6},
  year = {2021},
  month = {Aug},
  publisher = {American Physical Society},
  doi = {10.1103/PhysRevLett.127.086401}
}

@article{Drummond2009,
  title = {Phase Diagram of the Low-Density Two-Dimensional Homogeneous Electron Gas},
  author = {Drummond, N. D. and Needs, R. J.},
  journal = {Phys. Rev. Lett.},
  volume = {102},
  issue = {12},
  pages = {126402},
  numpages = {4},
  year = {2009},
  month = {Mar},
  publisher = {American Physical Society},
  doi = {10.1103/PhysRevLett.102.126402}
}

@article{Dornheim2018,
  title = {Ab initio Path Integral Monte Carlo Results for the Dynamic Structure Factor of Correlated Electrons: From the Electron Liquid to Warm Dense Matter},
  author = {Dornheim, T. and Groth, S. and Vorberger, J. and Bonitz, M.},
  journal = {Phys. Rev. Lett.},
  volume = {121},
  issue = {25},
  pages = {255001},
  numpages = {8},
  year = {2018},
  month = {Dec},
  publisher = {American Physical Society},
  doi = {10.1103/PhysRevLett.121.255001}
}

@article{Spink2013,
  title = {Quantum Monte Carlo study of the three-dimensional spin-polarized homogeneous electron gas},
  author = {Spink, G. G. and Needs, R. J. and Drummond, N. D.},
  journal = {Phys. Rev. B},
  volume = {88},
  issue = {8},
  pages = {085121},
  numpages = {9},
  year = {2013},
  month = {Aug},
  publisher = {American Physical Society},
  doi = {10.1103/PhysRevB.88.085121}
}

@article{Dornheim2018b,
title = {The uniform electron gas at warm dense matter conditions},
journal = {Phys. Reps.},
volume = {744},
pages = {1-86},
year = {2018},
issn = {0370-1573},
doi = {https://doi.org/10.1016/j.physrep.2018.04.001},
author = {T. Dornheim and S. Groth and M. Bonitz}
}

@article{Tanatar1989,
  title = {Ground state of the two-dimensional electron gas},
  author = {Tanatar, B. and Ceperley, D. M.},
  journal = {Phys. Rev. B},
  volume = {39},
  issue = {8},
  pages = {5005--5016},
  numpages = {0},
  year = {1989},
  month = {Mar},
  publisher = {American Physical Society},
  doi = {10.1103/PhysRevB.39.5005}
}

@article{Kwon1996,
    title = {Transient-estimate Monte Carlo in the two-dimensional electron gas},
    author = {Kwon, Y. and Ceperley, D. M. and Martin, R. M. M.},
    year = {1996},
    journal = {Phys. Rev. B},
    volume = {53},
    publisher = {American Physical Society},
    doi = {10.1103/PhysRevB.53.7376},
    pages = {7376--7382}
}

@article{Bockrath1999,
   author = {Bockrath, M. and Cobden, D. H. and Lu, J. and Rinzler, A. G. and Smalley, R. E. and Balents, L. and McEuen, P. L.},
   title = {Luttinger-liquid behaviour in carbon nanotubes},
   journal = {Nature},
   volume = {397},
   number = {6720},
   pages = {598–601},
   ISSN = {1476-4687},
   DOI = {10.1038/17569},
   year = {1999},
   type = {Journal Article}
}

@article{Ishii2003,
   author = {Ishii, H. and Kataura, H. and Shiozawa, H. and Yoshioka, H. and Otsubo, H. and Takayama, Y. and Miyahara, T. and Suzuki, S. and Achiba, Y. and Nakatake, M. and Narimura, T. and Higashiguchi, M. and Shimada, K. and Namatame, H. and Taniguchi, M.},
   title = {{Direct observation of Tomonaga–Luttinger-liquid state in carbon nanotubes at low temperatures}},
   journal = {Nature},
   volume = {426},
   number = {6966},
   pages = {540–544},
   ISSN = {1476-4687},
   DOI = {10.1038/nature02074},
   url = {https://doi.org/10.1038/nature02074},
   year = {2003},
   type = {Journal Article}
}

@article{Chang2003,
  title = {{Chiral Luttinger liquids at the fractional quantum Hall edge}},
  author = {Chang, A. M.},
  journal = {Rev. Mod. Phys.},
  volume = {75},
  issue = {4},
  pages = {1449--1505},
  numpages = {0},
  year = {2003},
  month = {Nov},
  publisher = {American Physical Society},
  doi = {10.1103/RevModPhys.75.1449}
}

@article{Schafer2008,
  title = {{New Model System for a One-Dimensional Electron Liquid: Self-Organized Atomic Gold Chains on Ge(001)}},
  author = {Sch\"afer, J. and Blumenstein, C. and Meyer, S. and Wisniewski, M. and Claessen, R.},
  journal = {Phys. Rev. Lett.},
  volume = {101},
  issue = {23},
  pages = {236802},
  numpages = {4},
  year = {2008},
  month = {Dec},
  publisher = {American Physical Society},
  doi = {10.1103/PhysRevLett.101.236802}
}

@article{Huang2001,
author = {Y. Huang  and X. Duan  and Y. Cui  and L. J. Lauhon  and K.-H. Kim  and C. M. Lieber },
title = {Logic Gates and Computation from Assembled Nanowire Building Blocks},
journal = {Science},
volume = {294},
number = {5545},
pages = {1313-1317},
year = {2001},
doi = {10.1126/science.1066192}
}

@article{Moritz2005,
  title = {{Confinement Induced Molecules in a 1D Fermi Gas}},
  author = {Moritz, Henning and St\"oferle, Thilo and G\"unter, Kenneth and K\"ohl, Michael and Esslinger, Tilman},
  journal = {Phys. Rev. Lett.},
  volume = {94},
  issue = {21},
  pages = {210401},
  numpages = {4},
  year = {2005},
  month = {Jun},
  publisher = {American Physical Society},
  doi = {10.1103/PhysRevLett.94.210401}
}

@book{Devreese1979,
  title        = {Highly Conducting One-Dimensional Solids},
  editor       = {J. T. Devreese and R. P. Evrard and V. E. Van Doren},
  year         = {1979},
  publisher    = {Plenum Press},
  address      = {New York},
  series       = {Physics of Solids and Liquids},
}

@article{Girardeau1960,
    author = {Girardeau, M.},
    title = {{Relationship between Systems of Impenetrable Bosons and Fermions in One Dimension}},
    journal = {J. Math. Phys.},
    volume = {1},
    number = {6},
    pages = {516-523},
    year = {1960},
    issn = {0022-2488},
    doi = {10.1063/1.1703687}
}

@article{Tomonaga1950,
    author = {Tomonaga, S.-I.},
    title = {Remarks on Bloch's Method of Sound Waves applied to Many-Fermion Problems},
    journal = {Progress of Theoretical Physics},
    volume = {5},
    number = {4},
    pages = {544-569},
    year = {1950},
    month = {07},
    issn = {0033-068X},
    doi = {10.1143/ptp/5.4.544}
}

@article{Luttinger1963,
    author = {Luttinger, J. M.},
    title = {An Exactly Soluble Model of a Many‐Fermion System},
    journal = {Journal of Mathematical Physics},
    volume = {4},
    number = {9},
    pages = {1154-1162},
    year = {1963},
    month = {09},
    issn = {0022-2488},
    doi = {10.1063/1.1704046}
}

@article{Haldane1981,
  title = {Effective Harmonic-Fluid Approach to Low-Energy Properties of One-Dimensional Quantum Fluids},
  author = {Haldane, F. D. M.},
  journal = {Phys. Rev. Lett.},
  volume = {47},
  issue = {25},
  pages = {1840--1843},
  numpages = {0},
  year = {1981},
  doi = {10.1103/PhysRevLett.47.1840},
  month = {Dec},
  publisher = {American Physical Society}
}

@article{Boninsegni2025,
   author = {Boninsegni, M.},
   title = {{Momentum distribution of one-dimensional $^3$He}},
   volume = {39},
   issue = {22},
   journal = {Int. J. Mod. Phys. B},
   pages = {2550208},
   doi  = {10.1142/S021797922550208X},
   year = {2025},
   type = {Journal Article}
}

@article{Casula2005,
   author = {Casula, M. and Senatore, G.},
   title = {{Charge and Spin Correlations of a One-Dimensional Electron Gas on the Continuum}},
   journal = {Chem. Phys. Chem.},
   volume = {6},
   number = {9},
   doi = {https://doi.org/10.1002/cphc.200500093},
   pages = {1902--1905},
   year = {2005},
   type = {Journal Article}
}

@article{Ashokan2018,
  title = {One-dimensional electron fluid at high density},
  author = {Ashokan, V. and Drummond, N. D. and Pathak, K. N.},
  journal = {Phys. Rev. B},
  volume = {98},
  issue = {12},
  pages = {125139},
  numpages = {9},
  year = {2018},
  month = {Sep},
  publisher = {American Physical Society},
  doi = {10.1103/PhysRevB.98.125139}
}

@article{Lee2011,
   author = {Lee, R. M. and Drummond, N. D.},
   title = {Ground-state properties of the one-dimensional electron liquid},
   journal = {Phys. Rev. B},
   volume = {83},
   number = {24},
   pages = {245114},
   note = {PRB},
   DOI = {10.1103/PhysRevB.83.245114},
   year = {2011},
   type = {Journal Article}
}

@article{Nemec2010,
  title = {{Diffusion Monte Carlo: Exponential scaling of computational cost for large systems}},
  author = {Nemec, N.},
  journal = {Phys. Rev. B},
  volume = {81},
  issue = {3},
  pages = {035119},
  numpages = {7},
  year = {2010},
  month = {Jan},
  publisher = {American Physical Society},
  doi = {10.1103/PhysRevB.81.035119}
}

@article{Boninsegni2012,
  title = {Population size bias in diffusion Monte Carlo},
  author = {Boninsegni, M. and Moroni, S.},
  journal = {Phys. Rev. E},
  volume = {86},
  issue = {5},
  pages = {056712},
  numpages = {7},
  year = {2012},
  month = {Nov},
  publisher = {American Physical Society},
  doi = {10.1103/PhysRevE.86.056712}
}

@article{Boninsegni2001,
  title = {Phase Separation in Mixtures of Hard Core Bosons},
  author = {Boninsegni, Massimo},
  journal = {Phys. Rev. Lett.},
  volume = {87},
  issue = {8},
  pages = {087201},
  numpages = {4},
  year = {2001},
  month = {Aug},
  publisher = {American Physical Society},
  doi = {10.1103/PhysRevLett.87.087201}
}

@article{DelMaestro2011,
   author = {Del Maestro, A. and Boninsegni, M. and Affleck, I.},
   title = {{$^4$He Luttinger Liquid in Nanopores}},
   journal = {Phys. Rev. Lett.},
   volume = {106},
   number = {10},
   pages = {105303–105303},
   DOI = {10.1103/PhysRevLett.106.105303},
   year = {2011},
   type = {Journal Article}
}

@article{Boninsegni2013,
   author = {Boninsegni, M.},
   title = {Ground State Phase Diagram of Parahydrogen in One Dimension},
   journal = {Phys. Rev. Lett.},
   volume = {111},
   number = {23},
   pages = {235303},
   DOI = {10.1103/PhysRevLett.111.235303},
   year = {2013},
   type = {Journal Article}
}

@article{Saunders1994,
title = {On the electrostatic potential in linear periodic polymers},
journal = {Comp. Phys. Comm.},
volume = {84},
number = {1},
pages = {156-172},
year = {1994},
issn = {0010-4655},
doi = {https://doi.org/10.1016/0010-4655(94)90209-7},
author = {V.R. Saunders and C. Freyria-Fava and R. Dovesi and C. Roetti}
}

@article{Mezzacapo2006b,
  title = {Structure, superfluidity, and quantum melting of hydrogen clusters},
  author = {Mezzacapo, Fabio and Boninsegni, Massimo},
  journal = {Phys. Rev. A},
  volume = {75},
  issue = {3},
  pages = {033201},
  numpages = {10},
  year = {2007},
  month = {Mar},
  publisher = {American Physical Society},
  doi = {10.1103/PhysRevA.75.033201}
}

@article{Mezzacapo2006,
  title = {Superfluidity and Quantum Melting of {\em p}-{H}$_2$ Clusters},
  author = {Mezzacapo, Fabio and Boninsegni, Massimo},
  journal = {Phys. Rev. Lett.},
  volume = {97},
  issue = {4},
  pages = {045301},
  numpages = {4},
  year = {2006},
  month = {Jul},
  publisher = {American Physical Society},
  doi = {10.1103/PhysRevLett.97.045301}
}

@article{Boninsegni2006b,
  title = {Worm algorithm and diagrammatic {Monte Carlo: A} new approach to continuous-space path integral Monte Carlo simulations},
  author = {Boninsegni, M. and Prokof'ev, N. V. and Svistunov, B. V.},
  journal = {Phys. Rev. E},
  volume = {74},
  issue = {3},
  pages = {036701},
  numpages = {16},
  year = {2006},
  month = {Sep},
  publisher = {American Physical Society},
  doi = {10.1103/PhysRevE.74.036701}
}

@article{Boninsegni2006,
  title = {Worm Algorithm for Continuous-Space Path Integral {Monte Carlo} Simulations},
  author = {Boninsegni, Massimo and Prokof'ev, Nikolay and Svistunov, Boris},
  journal = {Phys. Rev. Lett.},
  volume = {96},
  issue = {7},
  pages = {070601},
  numpages = {4},
  year = {2006},
  month = {Feb},
  publisher = {American Physical Society},
  doi = {10.1103/PhysRevLett.96.070601},
  url = {https://link.aps.org/doi/10.1103/PhysRevLett.96.070601}
}

@article{Jiang2001,
   author = {Jang, S. and Jang, S. and Voth, G. A.},
   title = {Applications of higher order composite factorization schemes in imaginary time path integral simulations},
   journal = {J. Chem. Phys.},
   volume = {115},
   number = {17},
   year = {2001},
   type = {Journal Article}
}

@article{Boninsegni2005,
   author = {Boninsegni, M.},
   title = {Permutation Sampling in Path Integral Monte Carlo},
   journal = {J. Low Temp. Phys.},
   volume = {141},
   number = {1-2},
   pages = {27–46},
   DOI = {10.1007/s10909-005-7513-0},
   year = {2005},
   type = {Journal Article}
}

@book{Giamarchi2003,
   author = {Giamarchi, Thierry},
   title = {Quantum Physics in One Dimension},
   publisher = {Oxford University Press},
   ISBN = {9780198525004},
   DOI = {10.1093/acprof:oso/9780198525004.001.0001},
   year = {2003},
   type = {Book}
}

@article{Schulz1993,
  title = {Wigner crystal in one dimension},
  author = {Schulz, H. J.},
  journal = {Phys. Rev. Lett.},
  volume = {71},
  issue = {12},
  pages = {1864--1867},
  numpages = {0},
  year = {1993},
  month = {Sep},
  publisher = {American Physical Society},
  doi = {10.1103/PhysRevLett.71.1864}
}

@article{Boninsegni2012b,
  title = {Role of Bose Statistics in Crystallization and Quantum Jamming},
  author = {Boninsegni, M. and Pollet, L. and Prokof'ev, N. and Svistunov, B.},
  journal = {Phys. Rev. Lett.},
  volume = {109},
  issue = {2},
  pages = {025302},
  numpages = {4},
  year = {2012},
  month = {Jul},
  publisher = {American Physical Society},
  doi = {10.1103/PhysRevLett.109.025302}
}

@article{Yukalov2005,
    title = {{Fermi-Bose mapping for one-dimensional Bose gases}},
    author = {Yukalow, V. I. and Girardeau, M. D.},
    journal = {Laser Phys. Lett.},
    volume = {2},
    doi = {10.1002/lapl.200510011},
    pages = {375--382},
    year = {2005}
}

\end{document}